\documentclass[twocolumn,pra,showpacs,superscriptaddress,amssymb,amsmath]{revtex4-1}
\usepackage{graphicx}
\usepackage{epstopdf}
\usepackage{bm}
\usepackage{hyperref}
\usepackage{wasysym}

\hypersetup{%
  pdfpagemode=None, 
  pdfstartpage=1,
  pdfstartview=FitH,
  pdfmenubar=true,
  pdftoolbar=true,
  colorlinks = true,
  linkcolor=blue,
  citecolor=blue,
  bookmarksopen=false
}

\begin{document}

\title{Chemical reactions of ultracold alkali-metal dimers in the lowest-energy $^3\Sigma$ state }

\author{Micha{\l} Tomza}
\address{Department of Chemistry, University of British Columbia, Vancouver, BC V6T 1Z1, Canada}
\address{Faculty of Chemistry, University of Warsaw, Pasteura 1, 02-093 Warsaw, Poland}
\author{Kirk W. Madison}
\address{Department of Physics, University of British Columbia, Vancouver, BC V6T 1Z1, Canada}
\author{Robert Moszynski}
\address{Faculty of Chemistry, University of Warsaw, Pasteura 1, 02-093 Warsaw, Poland}
\author{Roman V. Krems}
\address{Department of Chemistry, University of British Columbia, Vancouver, BC V6T 1Z1, Canada}

\date{\today}

\begin{abstract}

We show that the interaction of polar alkali dimers in the quintet spin state leads to the formation of a deeply bound reaction complex. The reaction complex can decompose adiabatically into homonuclear alkali dimers (for all molecules except KRb) and into alkali trimers (for all molecules). We show that there are no barriers for these chemical reactions. This means that all alkali dimers in the $a^3\Sigma^+$ state are chemically unstable at ultracold temperature, and the use of an optical lattice to segregate the molecules and suppress losses may be necessary. In addition, we calculate the minimum energy path for the chemical reactions of alkali hydrides. We find that the reaction of two molecules is accelerated by a strong attraction between the alkali atoms, leading to a barrierless process that produces hydrogen atoms with large kinetic energy.  We discuss the unique features of the chemical reactions of ultracold alkali dimers in the $a^3\Sigma^+$ electronic state.

\end{abstract}


\maketitle

The creation of ultracold, deeply bound dimers from laser cooled alkali-metal atoms can be achieved by photoassociation or by magnetoassociation followed by coherent transfer to a lower energy state by stimulated Raman adiabatic passage \cite{njp-review,NiScience08}.  The interaction of two ground-state alkali atoms gives rise to two molecular states: $X^1\Sigma^+$ and $a^3\Sigma^+$. The majority of experiments thus far have focused on the association of alkali atoms into the $X^1\Sigma^+$ state.  Fuelled by the promise of exciting applications \cite{Doyle04,njp-review}, the main goal of these experiments is to produce heteronuclear (polar) alkali dimers in the ro-vibrational ground state. The creation of polar alkali dimers in the ro-vibrational ground state of the $a^3\Sigma^+$ electronic state{~\cite{NiScience08}},
is currently emerging as another important research goal. Heteronuclear molecules in the $a^3\Sigma^+$ state offer both the electric and magnetic dipole moments. This can be exploited for a variety of novel applications \cite{MicheliNatPhys06,njp-review,pccp2008}.  However, alkali dimers in the $a^3\Sigma^+$ state may undergo inelastic collisions and chemical reactions necessitating the use of an optical lattice to segregate the molecules and suppress losses~\cite{PhysRevLett.108.080405}.

For alkali dimers AB$(a^3\Sigma^+)$ in the ground ro-vibrational state, the following reaction processes may lead to collisional losses:  
\begin{eqnarray}
{\rm AB}(a^3\Sigma^+)  + {\rm AB}(a^3\Sigma^+) &\rightarrow& {\rm A}_2(a^3\Sigma^+) + {\rm B}_2(a^3\Sigma^+)
\label{1}
\\
{\rm AB}(a^3\Sigma^+)  + {\rm AB}(a^3\Sigma^+) &\rightarrow& {\rm A}_2{\rm B} + {\rm B}
\label{2}
\\
{\rm AB}(a^3\Sigma^+)  + {\rm AB}(a^3\Sigma^+) &\rightarrow& {\rm A}_2(X^1\Sigma^+) + {\rm B}_2(T)
\label{3}
\\
{\rm AB}(a^3\Sigma^+)  + {\rm AB}(a^3\Sigma^+) &\rightarrow& {\rm AB}(X^1\Sigma^+) + {\rm AB}(T),
\label{4}
\end{eqnarray}
where $T $ is either $X^1\Sigma^+$ or $a^3\Sigma^+$. 
Reactions (\ref{3}) and (\ref{4}) can potentially be suppressed by confining AB($a^3\Sigma^+$) molecules in a magnetic trap. 
Magnetic trapping aligns the electron spin of molecules along the magnetic field axis, which restricts the 
total electron spin of the AB($a^3\Sigma^+$)  - AB($a^3\Sigma^+$) collision complex to the maximum value ${\cal S}=2$. 
Reactions (\ref{3}) and (\ref{4}) involve transitions to lower spin states mediated by non-adiabatic spin-dependent couplings \cite{gerrit}.  These couplings are induced by the long-range magnetic dipole - dipole interaction ($V_{\rm d-d}$) and the spin-dependent fine structure interactions effective at short intermolecular separations. The effect of $V_{\rm d-d}$ can generally be ignored \cite{gerrit}. The effect of the short-range couplings depends on the topology of the potential energy surface of the AB($a^3\Sigma^+$)  - AB($a^3\Sigma^+$) complex in the ${\cal S}=2$ state. The probability of reactions (\ref{1}) and (\ref{2}) is also determined by the AB($a^3\Sigma^+$)  - AB($a^3\Sigma^+$)  interaction surfaces. 

In the present work, we calculate the potential energy for the binary interactions of polar alkali dimers AB($a^3\Sigma^+$)  in the 
${\cal S}=2$ state of the two-molecule complex. The main goal is to explore the possibility of reaction barriers that would prevent molecules from reaching the short-range interaction region. It is known from previous calculations \cite{GutowskiJCP00,soldan,SoldanPRA08,SoldanPRA10} that the potential energy of alkali trimers is dominated by non-additive interactions. The same should be expected for the interaction of four alkali atoms. However, unlike in the atom - diatom case, reactions (\ref{1}) - (\ref{4}) involve the dissociation of two molecular bonds. 
The dissociation energy of these bonds may be expected to give rise to reaction barriers. We find no such barriers, meaning that reaction (\ref{1}), if energetically allowed, and reaction (\ref{2}) should be very fast at ultralow temperatures. Our calculations show that the non-additive three- and four-body interactions are much stronger than the binding energy of alkali dimers in the $a^3\Sigma^+$ state.

{
The potential energy surfaces reported here are calculated using the spin restricted open-shell coupled cluster method with the single, double, and noniterative triple excitations (RCCSD(T)) method. The Li and Na atoms were described with the augmented core-valence correlation consistent polarized Valence Triple-$\zeta$ atomic basis sets (aug-cc-pCVTZ)~\cite{DunningJCP89} and the H atom with the augmented correlation consistent polarized Valence Quadruple-$\zeta$ atomic basis sets (aug-cc-pVQZ).
The relativistic effects in the heavier alkali atoms were accounted for with the fully relativistic small-core energy consistent pseudopotentials MDF28 for Rb and MDF46 for Cs and the corresponding basis sets $[13s10p5d3f]/(8s7p5d3f)$ and $[12s11p6d4f]/(8s8p6d4f)$~\cite{LimJCP05}.  
The basis set superposition error was eliminated by using the counterpoise
correction of Boys and Bernardi~\cite{BSSE}.} All electronic structure calculations were performed with the MOLPRO package of {\it ab initio} programs~\cite{Molpro}.

\begin{figure}[t!]
\begin{center}
\includegraphics[width=0.99\columnwidth]{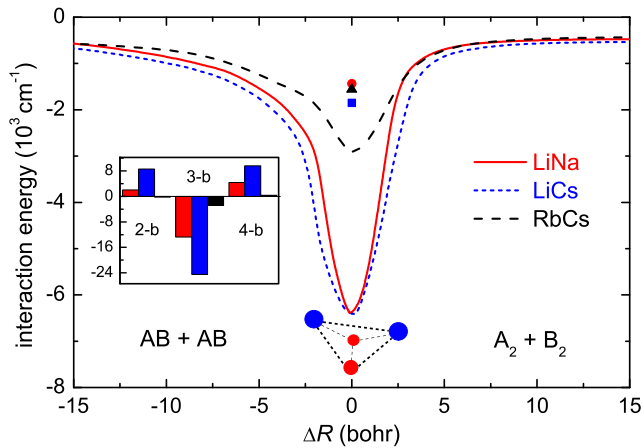}
\end{center}
\caption{(Color online) The minimum energy path of the adiabatic reaction for the LiCs-LiCs, LiNa-LiNa, and RbCs-RbCs reaction complexes in the quintet spin state from the optimized geometry calculations.
$\Delta R=(R_{AB}+R_{AB})/2-(R_{AA}+R_{BB})/2$, where $R_{AB}$ is the separation between the atoms $A$ and $B$. 
{The interaction energy equal to zero corresponds with all atoms dissociated.}
The symbols show the most negative values of the potential energy that can be obtained by adding binding energies of the dimers: $\CIRCLE$ - Li$_2$Na$_2$, $\blacktriangle$ - Rb$_2$Cs$_2$, $\blacksquare$ - Li$_2$Cs$_2$.
{ The inset shows the decomposition of the interaction energy for the reaction complexes at the minimum energy geometry into 2-, 3-, and 4-body contributions.}}
\label{fig:AB}
\end{figure}
\begin{figure}[t!]
\begin{center}
\includegraphics[width=0.99\columnwidth]{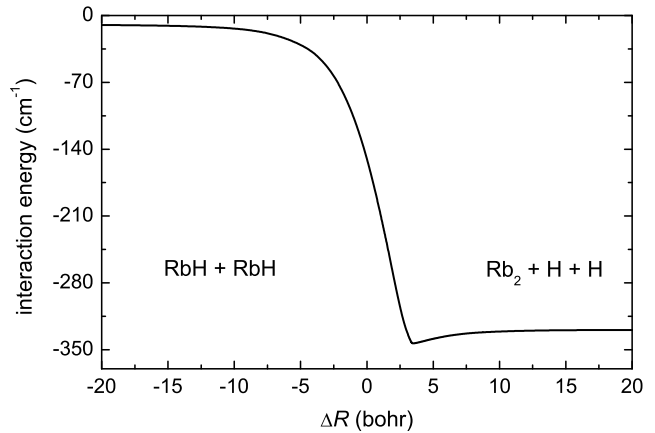}
\end{center}
\caption{The minimum energy path of the adiabatic reaction RbH + RbH $\to$ Rb$_2$ + H + H in the quintet spin state preserving the total electron spin from the optimized geometry calculations.
$\Delta R=R_{A\mathrm{H}}-R_{AA}$, where $R_{A\mathrm{H}}$ is the separation between the atoms $A$ and H.}
\label{fig:RbH}
\end{figure}

In order to prove the absence of reaction barriers in reactions (\ref{1}) and (\ref{2}), we calculated the potential energy of the four-atom complex along the minimum energy path of reaction (\ref{1}). The calculations were performed in two steps. 
First,  the minimum energy path was found by optimizing the geometries of the reaction complexes using the spin restricted open-shell coupled cluster method including single and double excitations (RCCSD), and basis sets { as described above but} truncated to $s,p$ and $d$ orbitals only. We defined the intermolecular coordinates $R_1$ and $R_2$ that specify the separation between the geometric centers of the heteronuclear molecules and the separations between the centers of the homonuclear molecules, respectively. The geometries were optimized at 20 values of $R_1$ and $R_2$ between the position of the global minimum and 40$\,$bohr by varying all other degrees of freedom. 
In the second step, the interaction energies for the optimized geometries were calculated using the more accurate RCCSD(T) method and the full basis sets. 
 For a few points we optimized the geometry with the full basis sets and the RCCSD(T) method and found that using the smaller basis set and the lower level of theory introduces negligible errors in the optimized geometry parameters but significantly underestimates the interaction energy.

Figure~\ref{fig:AB} presents the results of the calculations for the reactive interactions of LiNa, LiCs and RbCs molecules.
{ These molecules represent three limiting cases of polar alkali metal dimers: the lightest and most compact molecule, the most polar and the heaviest.} 
The four-body reactions are clearly barrierless and proceed through the formation of a stable reaction complex corresponding to the deep global minimum of the interaction potential surface. 
The reaction complex has a tetrahedral geometry as shown in Fig.~\ref{fig:AB}. The deep minimum of the potential energy is the manifestation of the non-additive forces in a four-body complex { (see instet of Fig.~\ref{fig:AB})}. 
Interestingly, the energy of the reaction paths for LiCs -- LiCs and LiNa -- LiNa are very similar, while that for RbCs -- RbCs is very different. This indicates that the non-additive interaction forces are largely mediated by the Li atoms. This is consistent with the results of Soldan {\it et al.}~\cite{soldan,SoldanPRA08,SoldanPRA10}. 

\begin{table*}[t!]
\caption{The dissociation energies $D_0$ (in cm$^{-1}$) for alkali metal dimers in the lowest triplet state $a^3\Sigma^+$. \label{tab:D0}}
\begin{ruledtabular}
\begin{tabular}{lrrrrr}
& Li & Na & K & Rb & Cs   \\
\hline
Li &  301.829(15)~\cite{LintonJMS99} & 211(10)   & 258.8(50)~\cite{TiemannPRA09}   & 257.6(40)~\cite{IvanovaJCP11}   & 287(10)\cite{StaanumPRA07} \\
Na  &   & 163.7(12)~\cite{HoJCP00}    & 196.48(10)~\cite{GerdesEPJD08}   &  193.365(50)~\cite{PashovPRA05} & 207.818(10)~\cite{DocenkoJPB06}  \\
K  &   &    & 244.523(50)~\cite{PashovEPJD08}  &  239.924(10)~\cite{PashovPRA07}  & 258.769(20)~\cite{FerberPRA09} \\
Rb  &   &    &  & 234.7641(30)~\cite{StraussPRA10}    & 252.316(30)~\cite{DocenkoPRA11}\\
Cs &   &    &   &    & 273.532(48)~\cite{XieJCP09} \\
\end{tabular}
\end{ruledtabular}
\end{table*}

While alkali metal dimers in the $a^3\Sigma^+$ state form molecules with multiple ro-vibrational states, the interaction of alkali atoms with hydrogen atoms in the $a^3\Sigma^+$ state gives rise to very shallow potential energy curves supporting only one bound state~\cite{rb-h}. 
Since the presence of multiple ro-vibrational states complicates
photoassociation of ultracold atoms, alkali hydrides in the $a^{3}\Sigma^+$
state appear to be attractive candidates for photoassociation experiments~\cite{CotePRA10}.
Such an experiment can be carried out, for example, by combining a slow, magnetically decelerated beam of hydrogen atoms with Rb atoms in a magneto-optical trap. 
In order to analyze the collisional stability of alkali hydrides thus formed, we extended the calculation of Figure~\ref{fig:AB} to compute the minimum energy path for the adiabatic reaction 2 RbH $\rightarrow$ Rb$_2$ + H + H, shown in Figure~\ref{fig:RbH}. 
Although there is no stable intermediate complex for this reaction, the reaction is barrierless. The strong attraction of the Rb atoms appears to pull the interacting molecules down a steep potential slope, resulting in the formation of free H atoms with large kinetic energy. Since most of the energy released as a result of the chemical reaction is carried away by the light hydrogen atoms, this may be used as an alternative way of creating ultracold Rb$_2$ molecules. 

While there are no reaction barriers to prevent reactions (\ref{1}) - (\ref{4}), some of the reaction channels may be energetically closed. The relative energies for the reactants and products for reactions (\ref{1}) and (\ref{2}) are summarized in Tables~\ref{tab:D0} - \ref{tab:dE_trimer}.
The dissociation energy of alkali dimers in the $a^3\Sigma^+$ state is known from spectroscopic measurements for all polar molecules except LiNa.  To complete the data, we calculated the binding energy of LiNa($a^3\Sigma^+$). For this calculation, we used the basis set aug-cc-pCVQZ~\cite{DunningJCP89} augmented by bond functions $(3s3p2d1f1g)$~\cite{midbond}. To estimate the error of the computations, we calculated the binding energies of both Li$_2$ and Na$_2$ molecules with the same method and basis sets. The results deviated from the experimental data by 3.5$\,$cm$^{-1}$.

Tables~\ref{tab:D0} - \ref{tab:dE_trimer} illustrate three important observations.  First, reaction (\ref{1}) is endothermic, and thus forbidden at ultralow temperatures, for KRb. Second, the change of energy in reaction (\ref{1}) is very small for any combination of alkali dimers. For example, the reaction KCs + KCs $\rightarrow$ K$_2$ + Cs$_2$ releases less than 1~cm$^{-1}$ of energy, whereas the reaction KRb + KRb $\rightarrow$ K$_2$ + Rb$_2$ requires an activation energy of about 0.6~cm$^{-1}$. 
This suggests that the former is bound to form diatomic molecules in the ground vibrational state and the latter can be stimulated by vibrational excitation of the reactants. Given that reaction (\ref{1}) combines polar species to form non-polar products, the probability of this reaction must be sensitive to external electric fields. Finally, Table~\ref{tab:dE_trimer} shows that reaction (\ref{2}) is exothermic for all combinations of molecules. In combination with the results of Figure~\ref{fig:AB}, this means that all alkali dimers in the $a^3\Sigma^+$ state are chemically reactive at ultralow temperatures. 
This is in contrast to alkali dimers in the ro-vibrational ground
state of the $X^1\Sigma^+$ electronic state for which the formation of trimers is always energetically forbidden making certain combinations of alkali dimers chemically stable~\cite{ZuchowskiPRA10}.

\begin{table}
\caption{The energy change $\Delta E$  (in cm$^{-1}$) for the reactions 2AB $\to$ A$_2$+B$_2$ of alkali dimer in the ro-vibrational ground state of the $a^3\Sigma^+$ electronic state. \label{tab:dE}}
\begin{ruledtabular}
\begin{tabular}{lrrrrr}
& Li & Na & K & Rb & Cs   \\
\hline
Li & 0 & -44(10) & -28.7(50) & -21.4(40) & -1.4(10)   \\
Na &   &       0 & -15.3(10) & -11.7(10) & -21.6(10) \\
K  &   &         &         0 & 0.561(60) & -0.52(10) \\
Rb &   &         &           &         0 & -3.66(10) \\
Cs &   &         &           &           & 0    
\end{tabular}
\end{ruledtabular}
\end{table}

\begin{table}
\caption{The energy change $\Delta E$  (in cm$^{-1}$) for the reactions 2AB $\to$ A$_2$B+B of alkali dimer in the ro-vibrational ground state of the $a^3\Sigma^+$ electronic state. The energies of the trimers were taken from Ref.~\cite{SoldanPRA10}.
\label{tab:dE_trimer}}
\begin{ruledtabular}
\begin{tabular}{lrrrrr}
$A \,\backslash \,B$ & Li & Na & K & Rb & Cs   \\
\hline
Li & -3647 & -2035 & -2280 & -2214 & -2609 \\
Na &  -953 &  -489 &  -587 &  -556 &  -685 \\
K  & -1316 &  -745 &  -803 &  -748 &  -858 \\
Rb & -1158 &  -643 &  -678 &  -620 &  -707 \\
Cs & -1579 &  -901 &  -907 &  -825 &  -897
\end{tabular}
\end{ruledtabular}
\end{table}

\begin{table}
\caption{The experimental equilibrium distance $R_e$, the value of the permanent dipole function $d_e$ at $R = R_e$, the permanent dipole moment of the molecule in the ro-vibrational ground state $d_0$, the rotational constant $B_0$ and the vibrational frequency $\omega_0$ of the alkali dimers in the $a^3\Sigma^+$ state. The reduced masses used in the calculations are for the most abundant isotopes. \label{tab:dip}}
\begin{ruledtabular}
\begin{tabular}{lrrrrr}
molecule & $R_e\,(a_0)$ & $d_e\,$(D) & $d_0\,$(D) & $B_0\,$(GHz) & $\omega_0\,$(cm$^{-1}$) \\
\hline
LiNa & 8.918 & 0.186 & 0.175 & 4.10 & 38.1 \\ 
LiK  & 9.433 & 0.321 & 0.312 & 3.35 & 40.6 \\ 
LiRb & 9.713 & 0.372 & 0.359 & 2.89 & 37.6 \\ 
LiCs & 9.916 & 0.475 & 0.462 & 2.70 & 41.1 \\ 
NaK  & 10.34 & 0.0283 & 0.0269 & 1.16 & 21.7 \\  
NaRb & 10.58 & 0.0592 & 0.0594 & 0.879 & 19.2 \\ 
NaCs & 10.86 & 0.0911 & 0.0914 & 0.772 & 18.7 \\ 
KRb  & 11.15 & 0.0508 & 0.0540 & 0.540 & 17.5 \\ 
KCs  & 11.44 & 0.101  & 0.101  & 0.454 & 16.4 \\ 
RbCs & 11.78 & 0.0348 & 0.0344 & 0.251 & 13.8 \\ 
LiH  & 11.28 & 0.0061 & 0.00051 & - & - \\
RbH  & 13.37 & 0.0061 & 0.00061 & - & - 
\end{tabular}
\end{ruledtabular}
\end{table}

\begin{figure}[t!]
\begin{center}
\includegraphics[width=\columnwidth]{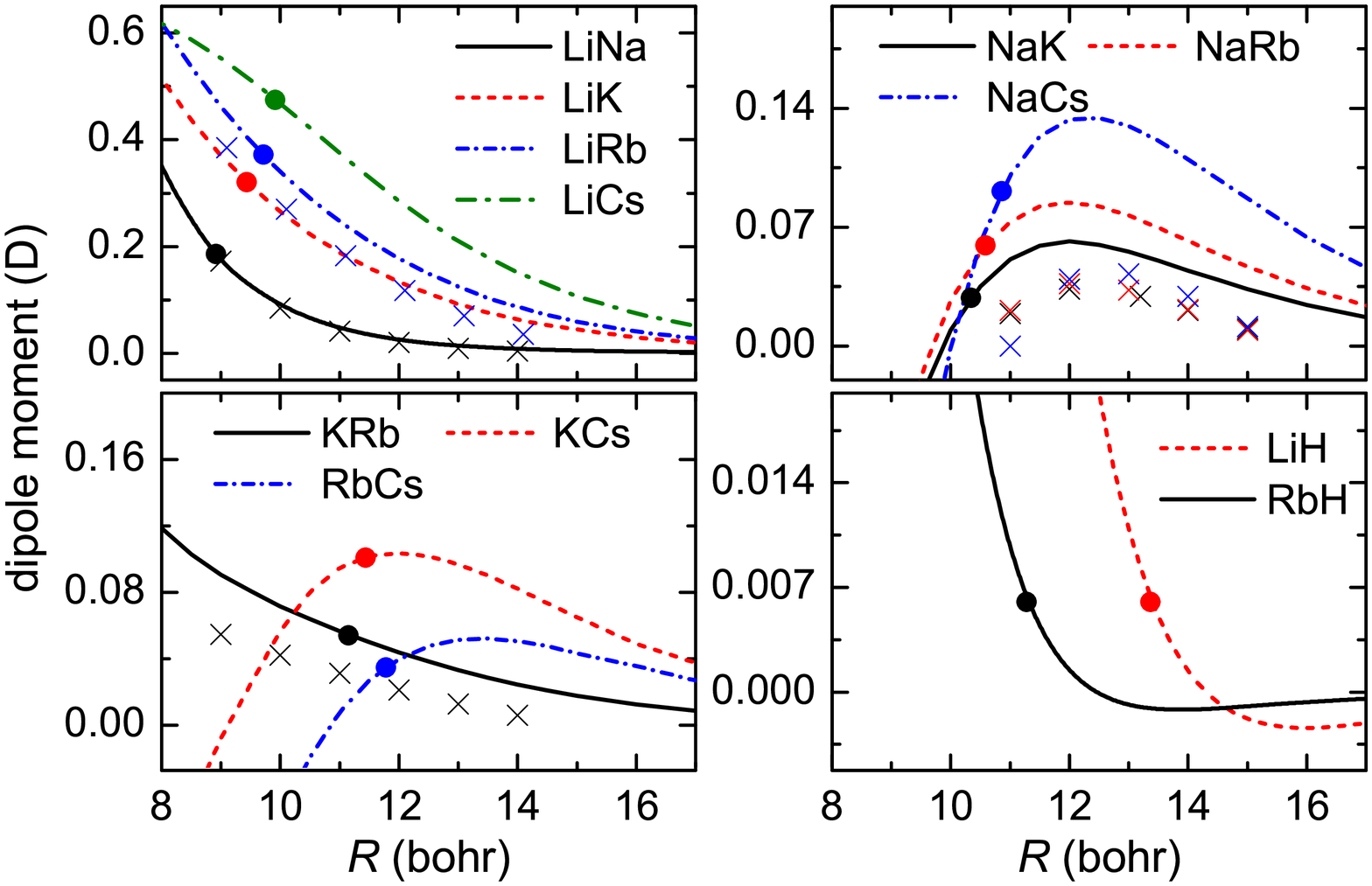}
\end{center}
\caption{(Color online) The permanent dipole moments of the heteronuclear alkali dimers and alkali metal hydrides in the $a^3\Sigma^+$ state. The {filled circles} indicate the value of the dipole moment at the equilibrium distance of the corresponding molecule {and the crosses are the results from Refs.~\cite{aymar,DeiglmayrJCP08}. The internuclear axis is oriented from the lighter atom to the heavier one.}}
\label{fig:dip}
\end{figure}

The results of Figures~\ref{fig:AB} and~\ref{fig:RbH} and Table~\ref{tab:dip} indicate that ultracold alkali dimers and alkali hydrides in the $a^3\Sigma^+$ state can be used for practical applications, only if protected from binary collisions by segregation in an optical lattice~\cite{Bloch05} or if confined  in a quasi-two-dimensional potential with their electric dipoles oriented {parallel and perpendicular to the plane of confinement~\cite{deMiranda}}.
All applications of molecules in optical lattices rely on the long-range dipole - dipole interactions. The magnitude of the permanent dipole moment is thus a figure of merit for experiments with molecules in optical lattices. Aymar and Dulieu presented a calculation of the potential energy curves and the dipole moments for all polar alkali dimers \cite{aymar,DeiglmayrJCP08}.
{ Their calculation treated alkali metals atoms as single-electron species with optimized pseudopotentials. 
The calculations of Refs.~\cite{aymar,DeiglmayrJCP08} included core polarization effects through effective terms and produced accurate results for the dipole moments of the alkali dimers in the $^1\Sigma$ state. However, the dipole moments of the molecules in the $^3\Sigma$ state have a smaller magnitude so they may be more sensitive to details of the calculations.}
We computed the dipole moments for the alkali dimers in the $a^3\Sigma^+$ state using the  RCCSD(T) approach with the aug-cc-pCVQZ basis for Na and Li, {the aug-cc-pVQZ basis for H,} and the small-core fully relativistic pseudopotentials MDFn~\cite{LimJCP05} and large basis sets for K ($[11s11p5d3f]$),  Rb ($[14s14p7d6f1g]$) and Cs ($[12s11p5d3f2g]$). These basis sets were optimized by calculating the energy of the electronic excitations in the individual atoms with the coupled cluster method~\cite{TomzaMP13,Kosc13}. In each case, the basis was augmented by the bond functions~\cite{midbond}. 
The results presented in Figure~\ref{fig:dip} and Table~\ref{tab:dip} { agree well with the calculations of Refs.~\cite{aymar,DeiglmayrJCP08} for light molecules containing Li but not for heavier molecules. 
Our results for RbCs differ from the previous calculations by a factor of ten, while agreeing within 5-10$\,\%$ with an independent calculation by Stolyarov~\cite{private-communication}. For KRb, our results agree 
to within $4\,\%$ with the experimental data~\cite{NiScience08} and the theoretical prediction by Kotochigova et al.~\cite{KotochigovaPRA03}, whereas the calculation in Ref.~\cite{aymar,DeiglmayrJCP08} underestimates the dipole moment for this molecule in the triplet state by $50\,\%$}.

In summary, we have shown that the interaction of heteronuclear alkali dimers in the lowest energy $a^3\Sigma^+$ state leads to the formation of a deeply bound reaction complex. The reaction complex -- that has a nearly symmetric tetrahedral configuration --  can decompose adiabatically into homonuclear alkali dimers (for all molecules except KRb) and into alkali trimers (for all molecules). There are no barriers for these chemical reactions. The absence of reaction barriers indicates the unique possibility to study interesting chemistry at ultralow temperatures. For example, measurements of the relative probabilities of reactions (\ref{1}) - (\ref{4}) in a magnetic trap  would reveal the role of the non-adiabatic spin-dependent interactions. 
The spin-dependent interactions are sensitive to external electric and magnetic fields \cite{timur1,timur2}, which can be used to manipulate the branching ratios with external fields. The relative energies of the reactants and products for reaction (\ref{1}) were found to be very close. This implies that the contribution of the reaction channel (\ref{1}) can be studied by measuring the chemical decay of molecules in different ro-vibrational states. This also suggests that the branching ratios of reactions (\ref{1}) and (\ref{2}) should be sensitive to external electric fields that can be used to shift the energy levels of the reactants by the amount of energy similar to the energy change in the chemical reaction~\cite{MeyerPRA10}.  

Our calculations illustrate the role of strong non-additive forces 
in four-body interactions of alkali atoms. We find that, as in the case of alkali trimers \cite{soldan,SoldanPRA08,SoldanPRA10}, these forces are much stronger for Li-containing molecules, making the minimum energy reaction paths of Li-containing molecules very similar. In addition, we calculated the minimum energy path for the chemical reactions of alkali hydrides. Since the binding energy of the alkali hydrides in the $a^3\Sigma^+$ state is very small, the reaction of two molecules is accelerated by a strong attraction between the alkali atoms, leading to a barrierless process that produces hydrogen atoms with large kinetic energy.  

Finally, we presented accurate calculations of the dipole moment functions for all alkali dimers as well as RbH and LiH in the $a^3\Sigma^+$ state. These calculations reveal that Li-containing alkali dimers have a substantial dipole moment in the ground ro-vibrational state, while the dipole moment of alkali hydrides LiH and RbH appears to be too small to be of practical use. 

{\it Acknowledgment} We acknowledge useful discussions with Professors Alexei Buchachenko, Andrei Stolyarov, and {Olivier Dulieu}.
The work reported here was initiated while the authors were visitors
at the Kavli Institute for Theoretical Physics, University of California at Santa Barbara
within the programme Fundamental Science and Applications of Ultracold
Polar Molecules.
Financial support from NSERC of Canada, the Polish Ministry of Science and Higher Education through the project No.~N~N204~215539, and the National Science Foundation grant No.~NSF~PHY11-25915 is gratefully acknowledged.
MT is supported by the project operated within the Foundation for Polish Science MPD Programme co-financed by the EU European Regional Development Fund.
RM thanks the Foundation for Polish Science for support within the MISTRZ programme.

\end{document}